\documentclass[prl,aps,twocolumn,floatfix,10pt]{revtex4}
\usepackage{graphicx,ifpdf}
\ifpdf 
\usepackage[pdftitle={PRL1996},pdfauthor={V. Nikolayev},pdftitle={PRL1996},colorlinks,pdftex]{hyperref} \fi
\begin{document}
\title{New hydrodynamic mechanism for drop coarsening}
\author{Vadim S. Nikolayev}
\altaffiliation{On leave from: Bogolyubov Institute for Theoretical Physics,
National Ukrainian Academy of Sciences, 252143, Kiev, Ukraine;
e-mail: vadim.nikolayev@cea.fr}
\author{Daniel Beysens}
\author{Patrick Guenoun}
\affiliation{Service de Physique de l'Etat Condens\'{e},
CE Saclay, F-91191 Gif-sur-Yvette Cedex,
France}
\date{9 March 1995}
\begin{abstract}
We discuss a new mechanism of drop coarsening due to coalescence
only, which describes the late stages of phase separation in fluids.
Depending on the volume fraction of the minority phase, we identify
two different regimes of growth, where the drops are interconnected
and their characteristic size grows linearly with time and where the
spherical drops are disconnected and the growth follows
(time)${}^{1/3}$.  The transition between the two regimes is sharp
and occurs at a well defined volume fraction of order 30\%.
\end{abstract}

 \maketitle

\begin{figure}\centering\includegraphics[width=8cm]{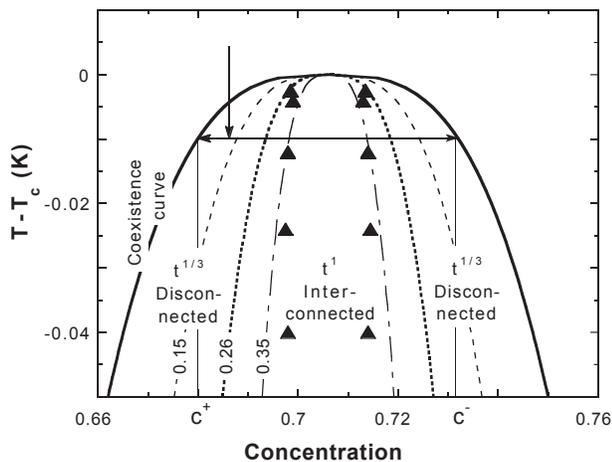}
\caption{Coexistence curve for a model two-phase system (density-matched cyclohexane--methanol) from Ref.
\cite{Jaya}. The dotted curve is the calculated boundary (see text) between the $t^{1/3}$ and $t^1$ growth
regions which corresponds to $\phi =0.26$. The triangles are the experimental data from Ref. 1.  The curves
corresponding to $\phi =0.15$ (random percolation limit) and $\phi =0.35$ (the value which gives the best fit to
the experimental data from Ref. 1) are presented also for comparison. The volume fraction of the minority phase
at the point $(c, T)$ can be calculated as $\phi=1/2-|c-c_c|/(c^--c^+)$ where $c_c$ (=0.707 for this case) is
the critical concentration.}\label{PhDiag} \end{figure}

In this Letter we concentrate on the kinetics of the late stages of
the phase separation. This subject has received considerable
attention recently \cite{Jaya}--\cite{Tanaka}. Most of the
experiments on growth kinetics have been performed near the critical
point of binary liquid mixtures (or simple fluids) because there the
critical slowing down allows the phenomenon to be observed during a
reasonable time.  After a thermal quench from the one-phase region
to the two-phase region of the phase diagram (Fig.  \ref{PhDiag}) the
domains of the new phases nucleate and grow.  It turns out that two
alternative regimes of coarsening are possible.  The first can be
observed when the volume fraction $\phi$ of the minority phase is
lower than some threshold \cite{Knob} and the domains of the
characteristic size $R$ grow according to the law $R\propto t^{1/3}$
($t$ is the time elapsed after the quench) as spherical drops.  The
second regime manifests itself when the quench is performed at high
volume fraction, the coarsening law is $R\propto t^1$ and the domains
grow as a complicated interconnected structure.  Recent experiments
\cite{Perrot} show that when $0.1<\phi<0.3$ the $t^{1/3}$ growth can
be explained by a mechanism of Brownian drop motion and coalescence
rather than the Lifshitz-Slyozov mechanism \cite{Lifsh} which holds
for $\phi\rightarrow 0$ and which we will not discuss here. We are
interested in the late stages of growth when phase boundaries are
already well developed and the concentrations of the phases are very
close to the equilibrium values at given temperature $T$ as defined
by the coexistence curve (Fig.~\ref{PhDiag}).  Then the drops grow
just because the system tends to minimize the total surface
separating the phases (i.e. due to coalescence) and $\phi$ no longer
depends on time.

{\bf Brownian coalescence.} The Brownian mechanism was considered
first by Smoluchowski \cite{Smo} for coagulation of colloids and was
then applied to phase separation by Binder and Stauffer \cite{Binder}
and Siggia \cite{Sig}.  According to this mechanism, the rate of
collisions per unit volume due to the Brownian motion of spherical
drops in the liquid of shear viscosity $\eta$ is \begin{equation}
N_B=16\pi DRn^2f(\phi), \label{NB} \end{equation} where $n$ is an
average number of drops per volume, $R$ is the average radius of the
drops and $D$ is the diffusion coefficient of the drops of the same
viscosity
\begin{equation}
D={k_BT\over 5\pi\eta R}\quad .
\end{equation}
The factor $f$ represents the correction which takes into account the
hydrodynamic interaction between the drops. It depends on the ratio
of the viscosities of the liquid inside and outside the drops and the
average distance between the drops and, therefore, on the volume
fraction $\phi$. This correction has been calculated in the dilute
limit ($\phi\rightarrow 0$) by Zhang and Davis \cite{Zhang}.  In
the proximity of the critical point we can assume the viscosities of
the phases to be equal and \begin{equation}
f(0)=0.56.\label{f0} \end{equation} According to the model, the drops
coalesce immediately after the collision. Coalescence is the only
reason for the decrease of the total number of the drops with the
rate  \begin{equation} {dn\over dt}=-N_B.\label{Brate} \end{equation}
With the relation \begin{equation} \phi ={4\over 3}\pi R^3n =
\mbox{const}, \label{rel} \end{equation} Eq.(\ref{Brate}) yields a
$R\propto t^{1/3}$ law. The further improvements (see \cite{Hay}
and refs. therein) of this model influence mainly the numerical factor
in (\ref{NB}) which is not important for the present considerations.

{\bf Hydrodynamic approaches.} The origin of the $R\propto t^1$
growth law, observed at high volume fraction where domains are
interconnected, is much less clear. By means of a dimensional analysis
Siggia \cite{Sig} has shown that hydrodynamics is needed to explain
the kinetics. It assumed the growth to be ruled by the Taylor
instability of the long tube of fluid, which breaks into separate
drops and associated the growth rate with the rate of the evolution
of the unstable fluctuations. This idea has been developed by San
Miguel {\it et al.} in \cite{SanMig}.  However, it was not clear how
this process was related to the growth.

Another approach has been considered by Kawasaki and Ohta \cite{Kaw}
who used a model of coupled equations of hydrodynamics and diffusion.
It was assumed that the growth is controlled by diffusion and the
hydrodynamic correction to this process was calculated. However, the
translational movement of the drops due to the pressure gradient
was not taken into account.  The motion of the liquid was supposed to
be induced by the concentration gradient only.  At the same time, it
is well known that the concentration variation does not enter the
equations of hydrodynamics of the liquid mixture in a first
approximation \cite{Land}. Moreover, it is evident that at high
volume fraction the coalescence process induced by the translational
motion of the drops becomes very important.

Recently, several groups \cite{puri,alex,valls,shino,appert} performed large
scale direct numerical simulations by using different approaches to
solve coupled equations of diffusion and hydrodynamics. Some
recovered the linear growth law  \cite{alex,shino}, whilst the
others were not able to reach the late stages of separation and
measured the transient values of the growth exponent (between 1/3 and
1). In spite of these efforts, the physical mechanism for the linear
growth has not been clarified. To our knowledge, the simulations
never showed two asymptotical laws depending on the volume fraction:
the exponent is either larger than 1/3 when accounting for
hydrodynamics or 1/3 for pure diffusion. Thus the precise threshold
in $\phi$ separating the $t^1$ and $t^{1/3}$ regimes is not predicted
either by any theory or by simulation.

\begin{figure}\centering\includegraphics[width=5cm]{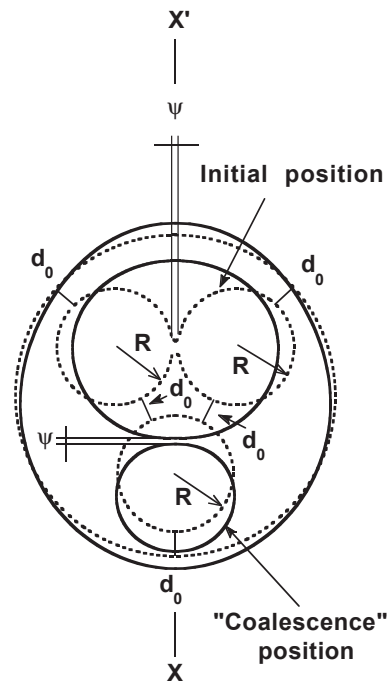}
\caption{The positions of the drop surfaces at the beginning (dotted line) and at the end (solid line) of the
simulation. $\psi$ is the coalescence distance, $d_0$ is the initial distance between the drops, $X'$--$X$ is
the axis of cylindrical symmetry. \label{Surf}} \end{figure}

{\bf Simulation of coalescence.} We show here that a $t^1$ growth can
originate from a coalescence mechanism whose limiting process between
two coalescences is not Brownian diffusion but rather flows induced
by previous coalescence.  We use the concept of ``coalescence-induced
coalescence'' as introduced by Tanaka \cite{Tanaka} who, however,
thought that induction by the hydrodynamic flow was not relevant,
having stated that coalescence takes place after the decay of the
flow. We consider here a coalescence process between two drops and
study numerically the generated flow and its influence on the third
neighboring drop. The fully deterministic hydrodynamic problem within the
 creeping flow approximation (which is well justified near the critical point
\cite{Sig}) was solved. The free boundary conditions were applied on
all drop interfaces whose motion  is driven by surface
tension. At each time step the velocity of each mesh point on the interface
contours
 was computed using a boundary integral approach
\cite{Pozdr}. When the new positions of the interfaces have been calculated, the procedure was continued
iteratively.
 The details of the solution will be
presented elsewhere \cite{Prep}.  We begin the simulation when coalescence starts between two drops of size $R$
(Fig.~\ref{Surf}), i.e.  when the drops of the minority phase approach to within a distance of coalescence
$\psi$ which corresponds to the interface thickness of the drop \cite{Sig}.  We choose for the simulation
$R=10\psi$. Then we place another drop at the distance $d_0$ from the composite drop (defined as the aggregate
of two coalescing drops) and envelope these two drops by a spherical shell to mimic the surrounding pattern of
tightly packed drops. Thus the distance between the drops and the shell has been chosen to be $d_0$ also. The
surface tension $\sigma$ is supposed to be the same for all the interfaces.  Unfortunately, due to the
prohibitively long computing times, we could not simulate the process of coalescence of two {\em spherical}
drops. Instead, we had to use a configuration with cylindrical symmetry with respect to the axis $X'$ -- $X$
(see Fig.~\ref{Surf}) which is expected to retain the main features associated with the spherical shape. In the
beginning of the simulation the composite drop looks more like a torus. The spherical shell approximation can be
justified by the fact that the main effect of the assembly of surrounding drops (as well as of the spherical
shell) is to confine the motion of the neighboring drops -- see \cite{Prep} for an advanced discussion. A setup
with only two drops without either a shell or surrounding drops would not enable a new coalescence even for the
smallest interdrop distances.  Though we cannot control quantitatively this approximation, it is the simplest
one which captures the main features of the process.

A first important result from the simulation is that the flow
generated by the first coalescence is able to generate {\em a second
coalescence} between the composite drop and the neighboring drop
(Fig.~\ref{Surf}). This means that the lubrication interaction with
the surrounding drops (with the shell in our model) make attract the
composite and the neighboring drops (note that the second coalescence
{\em does not} take place between this neighboring drop and the
shell).  This  leads to the formation of a new elongated
droplet. When the drops are close enough to each other the composite
drop will have no time to relax to a spherical shape since a new
coalescence takes place before relaxation. An interconnected pattern
naturally follows.  In contrast, when the drops are far from each
other ($d_0/R\gg 1$), the second coalescence will never take place.
The droplets take a spherical shape and the liquid motion stops. It
is also clear that if $d_0/R < l_G$, which we call ``geometric
coalescence limit'', coalescence necessary occurs due to geometrical
constraint. We find $l_G\approx 0.484$ \cite{Prep}, \cite{l_G}.

\begin{figure}
\centering\includegraphics[width=8cm]{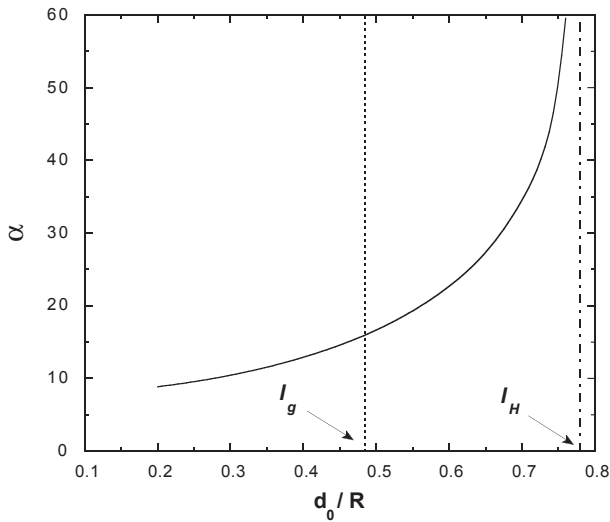}\caption{The simulation data on the dependence of the reduced
coalescence time $\alpha$ on $d_0/R$. The vertical lines show the geometric and hydrodynamic coalescence
limits.\label{Time}} \end{figure}

The second important result is that the coalescence takes place also
for $l_G<d_0/R<l_H$ where $l_H$ is a value which we call
``hydrodynamic coalescence limit''. It is defined as a reduced
initial distance where the time between two coalescences ($t_c$)
becomes infinite . Since there is only one length scale ($R$) in the
problem, $t_c$ can be written in the scaled form \begin{equation}
t_c=\alpha\eta R/\sigma, \label{TcH}
\end{equation}
where $\alpha$ is a reduced coalescence time which depends on $d_0/R$
only (Fig.~\ref{Time}).

The quantity $d_0/R$ is related to the volume fraction of the
drops (minority phase) $\phi$:
\begin{equation}
\phi =b[1+d_0/(2R)]^{-3}.\label{phi}
\end{equation}
The constant $b$ depends on the space arrangement of the drops. The
hydrodynamic interaction between them is always repulsive due to the
lubrication force. Thus they tend to be as far from each other as
possible.  Moreover, experiment shows a liquid-like order for the
drop positions. Such a correlation explains why the drops do not
percolate \cite{Jaya} when the volume fraction $\phi$ reaches the
{\it random} percolation limit $(\phi\approx 0.15)$.

Since no quantitative information is available to determine $b$, we
calculate its upper and lower bounds. Ideally, we can assume that the
drops are arranged into a regular lattice, with the vertices as far
from each other as possible. This is the face centered cubic lattice
where $b=\pi/3\sqrt{2}\approx 0.74$ and which corresponds to the
fully ordered structure. We can also consider as a lower bound the
random close packing arrangement for spheres of radius $R+d_0/2$.
This corresponds to the absence of a short-range order \cite{RP} and
implies $b\approx 0.64$. We note that the value of $b$ is not very
sensitive to the particular space arrangement. In the following we
adopt the median value $b=0.69$.

{\bf Generalization of the hydrodynamic model.} Now we can
generalize the above hydrodynamic mechanism for an arbitrary shape
of the drops. The self-similarity of the growth implies the following
relation for the characteristic sizes of the drops between $i$-th and
$(i+1)$-th coalescences: $R^{(i+1)}=\beta R^{(i)}$, where $\beta$ is
a universal shape factor, which depends on $\phi$ only. We can
rewrite also the Eq. (\ref{TcH}) for the time between the
coalescences in the form $t_c^{(i)}=\alpha(\phi)\eta R^{(i)}/\sigma$,
where $\alpha(\phi)$ is also the universal function. The last
expression conforms to the scaling assumption which implies the
independence of $t_c^{(i)}$ on the second length scale $\psi$. Then
after $n$ coalescences
$$R=\beta^nR^{(0)},\quad t=\sum_{i=0}^{n-1} t_c^{(i)}$$ and
\begin{equation}
R=R^{(0)}+{\beta-1\over\alpha}\cdot{\sigma\over\eta}t, \label{Rt}
\end{equation}
where $R^{(0)}$ is the initial size of the drop.  Noting that $\beta=2^{1/3}$ for the spheres and $\beta\gtrsim
1$ for the long tubes, we can take $\beta\sim 1.1$ for the estimate. Since $\alpha\sim 10$ for $\phi=0.5$, as it
follows from Fig.~\ref{Time} and Eq.(\ref{phi}), we obtain $R\sim 0.01\sigma/\eta\,t$, which compares well with
the experiment \cite{Rate}, which gives 0.03 for the numerical factor.

{\bf Competition between two mechanisms.} Using now Eq. (\ref{phi})
we can relate $l_H$ to a volume fraction $\phi_H$. It is clear that
the described hydrodynamic mechanism works only when $\phi>\phi_H$
while the Brownian coalescence takes place in the whole range of
$\phi$. Below we shall consider the regime for which $\phi>\phi_H$ in
order to obtain the position of the boundary between $t^{1/3}$ and
$t^1$ regions on the phase diagram.

Taking into account the competition between the two mechanisms, we
consider the relation \begin{equation} {dn\over
dt}=-(N_B+N_H),\label{BHrate} \end{equation} instead of
(\ref{Brate}), where $N_H$ is the rate of the coalescences due to the
hydrodynamics which can be calculated by using Eq.(\ref{Rt}) and
the relation between $n$ and $R$ (Eq.(5)).  The latter, however, depends on
the shape of the drop. We assume that in the early stages the drops
are spherical and we use Eq.(\ref{rel}).  It should be noted that the
Eq.(\ref{BHrate}) rewritten for the scaled wavenumber exactly coincides
 with the semi-empirical equation suggested by Furukawa
\cite{Furukawa}.

The Brownian mechanism dominates when $N_B>N_H$. In the vicinity
of the critical point one can use two-scale-factor universality
\cite{2scale} expression $\sigma=k_BT/\gamma\xi^2$ where $\xi$ is the
correlation length in the two-phase region, $\gamma\approx 0.39$ is a
universal constant. Then we reduce the last inequality to
\begin{eqnarray} R^2/\xi^2<G(\phi), \label{Bound} \\
\mbox{where}\quad G(\phi)=C\phi f(\phi)\alpha(\phi)\quad
\mbox{and}\quad C={\gamma\over 5\pi(\beta-1)}.\nonumber
\end{eqnarray}
We do not know much
about the function $f(\phi)$ which has been discussed in
Eq.(\ref{NB}). However, it is unphysical to assume that it exhibits
singularities or a steep behavior. We shall use the constant
value (\ref{f0}) in the following. We recall that all our
considerations are valid only when the drop interfaces have already
formed (late stages of growth), which means that the initial radius
of a drop cannot be less than the interface thickness, i. e.  $\approx
4\xi$
\cite{Thick}. It follows readily from the inequality (\ref{Bound})
that growth would obey the law $R\propto t^{1/3}$ when
\begin{equation} G(\phi)>16.\label{ineq} \end{equation} Now we aim to
estimate the function $G(\phi)$ by using for $\alpha(\phi)$ the
calculated function in Fig.~\ref{Time} along with Eq.(\ref{phi}). It
turns out that the function $G(\phi)$ well fits the power law
\begin{equation} G(\phi)\propto
(\phi-\phi_H)^{-\delta} \label{Fit} \end{equation}
for $\phi>\phi_H$ where $\phi_H\approx 0.26$, $\delta\approx
0.33$ and the divergency comes from $\alpha(\phi)$. From
(\ref{Fit}), it is easy to deduce that (\ref{ineq}) is valid when
$0<\phi-\phi_H\lesssim 10^{-6}$. In practice, this means that for all
$\phi>\phi_H$, the hydrodynamic mechanism {\em only} will determine
the growth from the very beginning of the drop coarsening.
Alternatively, for $\phi<\phi_H$, the drops will grow according to
the Brownian mechanism only. This explains the sharp transition in
the kinetics ($t^1\rightarrow t^{1/3}$) which is controlled by the
volume fraction of the minority phase as observed in \cite{Jaya} and
\cite{Perrot}. The curve which corresponds to the threshold value
$\phi = 0.26$ is plotted in Fig.~\ref{PhDiag}. It shows a reasonable
agreement with the experimental data in spite of our very particular
choice of the form and arrangement of the drops.

It should be mentioned that our model can be applied to any system
where the growth is due to the coalescence of liquid drops inside
another fluid (phase separation, coagulation, etc.).

\acknowledgments
One of the authors (V.~N.) would like to thank the collaborators of
SPEC Saclay for their kind hospitality and Minist\`{e}re de
l'Enseignement Sup\'{e}rieur et de la Recherche of France for the
financial support.


\begin{references}
\bibitem{l_G} The shell diameter $D_s$ at the initial moment can be
defined according to Fig.~2 as a function of $R$, $\psi$ and $d_0$.
If there is no coalescence, the composite drop  eventually becomes
a sphere with a diameter $D_c$ which can be deduced out of its initial
volume, because the latter is conserved during the evolution. The $l_G$
value is
determined by the condition $D_s=D_c+2R+3\psi$ which means that
two drops of the final spherical shape must fit inside the spherical
shell with the gaps of the width $\psi$.
\end{references}
\end{document}